\documentclass[manuscript]{emulateapj}

\shorttitle{Near-infrared spectroscopy of EX Lupi in outburst}
\shortauthors{K\'osp\'al et al.}

\begin{document}


\title{Near-infrared spectroscopy of EX Lupi in outburst}

\author{\'A. K\'osp\'al\altaffilmark{1},
P. \'Abrah\'am\altaffilmark{2},
M. Goto\altaffilmark{3},
Zs. Reg\'aly\altaffilmark{2},
C.P. Dullemond\altaffilmark{3},
Th. Henning\altaffilmark{3},
A. Juh\'asz\altaffilmark{1},
A. Sicilia-Aguilar\altaffilmark{3},
M. van den Ancker\altaffilmark{4}}
\email{kospal@strw.leidenuniv.nl}

\altaffiltext{1}{Leiden Observatory, Leiden University, P.O. Box 9513,
  NL-2300 RA Leiden, The Netherlands}

\altaffiltext{2}{Konkoly Observatory of the Hungarian Academy of
  Sciences, P.O. Box 67, H-1525 Budapest, Hungary}

\altaffiltext{3}{Max Planck Institute for Astronomy, K\"onigstuhl 17,
  D-69117 Heidelberg, Germany}

\altaffiltext{4}{ESO, Karl-Schwarzschild-Stra\ss{}e 2, D-85748
  Garching bei M\"unchen, Germany}


\begin{abstract}
  EX\,Lup is the prototype of the EXor class of young eruptive stars:
  objects showing repetitive brightenings due to increased accretion
  from the circumstellar disk to the star. In this paper, we report on
  medium-resolution near-infrared spectroscopy of EX\,Lup taken during
  its extreme outburst in 2008, as well as numerical modeling with the
  aim of determining the physical conditions around the star.  We
  detect emission lines from atomic hydrogen, helium, and metals, as
  well as first overtone bandhead emission from carbon monoxide. Our
  results indicate that the emission lines are originating from gas
  located in a dust-free region within $\approx$0.2\,AU of the
  star. The profile of the CO bandhead indicates that the CO gas has a
  temperature of 2500\,K, and is located in the inner edge of the disk
  or in the outer parts of funnel flows. The atomic metals are
  probably co-located with the CO. Some metallic lines are
  fluorescently excited, suggesting direct exposure to ultraviolet
  photons. The Brackett series indicates emission from hot
  (10\,000\,K) and optically thin gas. The hydrogen lines display a
  strong spectro-astrometric signal, suggesting that the hydrogen
  emission is probably not coming from an equatorial boundary layer; a
  funnel flow or disk wind origin is more likely. This picture is
  broadly consistent with the standard magnetospheric accretion model
  usually assumed for normally accreting T\,Tauri stars. Our results
  also set constraints on the eruption mechanism, supporting a model
  where material piles up around the corotation radius and
  episodically falls onto the star.
\end{abstract}



\section{Introduction}

Young eruptive stars form a special class of pre-main sequence
objects. The class is defined by unpredictable brightenings of up to
5\,mag at optical wavelengths, and it is believed that temporarily
increased mass accretion from the circumstellar disk onto the star is
responsible for the flare-ups \citep{hk96}. In spite of their
relatively small number (around two dozen stars), young eruptive stars
are key objects in understanding early stellar evolution. On the one
hand, the mass inflow during outbursts contributes to the build-up of
the final stellar mass, which would otherwise be difficult to explain
considering the observed typical rates of mass accretion \citep[less
  than $\approx$10$^{-6}$\,M$_{\odot}\rm
  yr^{-1}$,][]{kenyon1994,evans2009}. On the other hand, the eruptions
affect the density, temperature, and chemical structure of the disk,
possibly also influencing the conditions of planetesimal and planet
formation in the terrestrial zone \citep{abraham2009}.

One of the two main groups of young eruptive stars is called EXors,
named after the prototype EX\,Lup. These objects produce smaller,
1-3\,mag flare-ups every few years for a period of a few weeks as well
as rare 5 mag extreme outbursts lasting several months (see e.g.~the
light curve of VY\,Tau in \citealt{herbig1977} or that of EX\,Lup in
\citealt{herbig2007}). Although accretion is the likely source of
power for the outbursts, fundamental aspects of the mass inflow
process in EXors require significant clarification.

For a normally accreting T\,Tauri star, the stellar magnetic field
truncates the accretion disk at a distance of a few stellar radii
above the stellar surface. From that point, the material is
magnetically channeled onto the stellar surface along funnel flows
\citep{bouvier2007}. During outburst, the accretion rate is enhanced
by a few orders of magnitude, and it may be the case that the magnetic
field can no longer truncate the inner regions of the accretion disk
and hence disk material is dumped onto the star along the equatorial
plane in a boundary layer \citep{popham1996}. It remains an open
question as to what path the accreting material follows in young
eruptive stars like EXors.

EX\,Lup \citep[d=155\,pc,][]{lombardi2008} produced its most extreme
eruption in 2008 \citep{jones2008,kospal2008b}. This made it the
center of attention and the subject of many different kinds of
observations. Our group conducted a coordinated multi-wavelength
observing campaign at a single epoch, as well as monitoring studies
during the outburst.

The optical spectroscopic data of this campaign, which will provide
insight into the accretion and wind processes, will be presented in
\citet{sicilia-aguilar2011}. In \citet{goto2011} we analyzed the CO
fundamental emission lines using high resolution spectra in the
thermal near-infrared and found that the emission comes from two
separate physical components: a narrow line component which is
constant in time, and a broad line component which is closely related
to the outburst and is decaying with time. In \citet{abraham2009} we
reported on on-going crystallization of silicate grains during the
outburst, and our monitoring of the crystalline silicate features
showed evidence for a fast radial transport of silicate crystals
\citep{juhasz2010}. In Juh\'asz et al.~we also modeled the whole
optical-to-mm spectral energy distribution of EX\,Lup in outburst and
found that most of the accretion luminosity is emitted as a single
temperature blackbody of 6500\,K.

In this paper we present the medium resolution near-infrared spectra
of our campaign, and use the observed emission lines to analyze the
location, kinematics, and energetics of warm atomic and molecular gas
present in the inner few tenths of AUs in the EX\,Lup system.


\section{Observations and data reduction}

\subsection{Near-infrared spectroscopy}

We observed EX\,Lup with SINFONI, an adaptive optics (AO) assisted
integral field spectrograph installed on the UT4 telescope of the VLT
\citep{eisenhauer2003, bonnet2004}. Measurements were taken in service
mode at three different epochs, on the nights 24/25, 28/29, and 30/31
July 2008, as part of the project 281.C-5031 (PI: M.~Goto). Three
dimensional spectra were taken with the J, H and K gratings using a
Hawaii 2RG detector. The dispersion of the selected mode was
0.145\,nm/pixel in the J band (1.10--1.45$\,\mu$m), 0.195\,nm/pixel in
the H band (1.45--1.85$\,\mu$m), and 0.245\,nm/pixel in the K band
(1.93--2.45$\,\mu$m). The spectral resolution was $\lambda /
\Delta\lambda\,\approx$ 2400, 4100, and 4400 in the J, H, and K band,
respectively. Spectra were obtained by rotating the instrument by
0$^{\circ}$, 180$^{\circ}$, 90$^{\circ}$, and 270$^{\circ}$, in order
to be able to better remove instrumental effects. Exposure time was
1\,min for each rotator angle, giving a total exposure time of 4\,min
per band per epoch. Beside EX\,Lup, calibration sources of spectral
type G2V were also observed in order to enable proper telluric
correction: HIP\,93685 and HIP\,79464 were observed in the J band on
25 July, HIP\,80982 was observed in all three bands on Jul 29, and
HIP\,78652 was observed also in all three bands on Jul 31.

\begin{figure}
\begin{center}
\includegraphics[angle=0,width=\columnwidth]{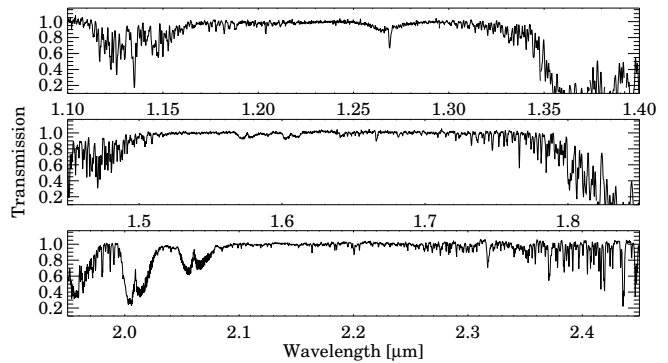}
\caption{Transmission of the atmosphere at Paranal Observatory on the
  night of 30/31 July 2008, in the J, H, and K bands, showing lines and
  bands of telluric absorption. \label{fig:calib}}
\end{center}
\end{figure}

We started the data reduction with product files provided by the
SINFONI pipeline as part of the service mode data delivery: product
code SCDJ (full, coadded science product cubes) for the science
target, and PCST (full, coadded standard star cube) for the telluric
standard. These files contain dark current-subtracted,
flatfield-corrected, sky-subtracted, wavelength-calibrated, co-added
data products in the form of 64\,pixel$\,{\times}\,$64\,pixel images
for each wavelength, with a spatial scale of 12.5\,mas/pixel and a
field of view of 0$\farcs$8$\,{\times}\,$0$\farcs$8. Further data
reduction was done using custom-written IDL scripts. For both the
science target and the telluric calibration stars, we first calculated
the centroid for each image in the data cubes. Then we extracted
spectra by using an aperture with a radius of 10 pixel and a sky
annulus between 20 and 25 pixel to calculate the flux of the star at
each wavelength. The spectrum of the telluric standard star contained
not only the telluric absorption lines, but intrinsic stellar
photospheric lines. To correct for the latter, we used the normalized
solar spectrum available at the SINFONI webpage\footnote{These spectra
  were created from data made available by the NSO/Kitt Peak
  Observatory, produced by NSF/NOAO.}. We multiplied it by a
blackbody, convolved it to the spectral resolution of SINFONI and
scaled it. By changing the temperature of the blackbody, the width of
the Gaussian convolution kernel, and the scaling factor, it was
possible to remove any intrinsic stellar features from the telluric
correction spectra. The obtained correction curves characterize the
transmission of the atmosphere as a function of wavelength
(Fig.~\ref{fig:calib}). Since for the first night no H and K
calibration observations were available, we used the average of the
curves obtained on the two other nights. Finally, the spectra of
EX\,Lup were divided by these correction curves, and the results were
normalized by fitting a third degree polynomial to the continuum.

Comparison of the spectra taken on the three different nights revealed
insignificant differences, thus we averaged them to increase the
signal-to-noise ratio (S:N). For most of the following analysis, we
used these averaged spectra, which are plotted in
Figs.~\ref{fig:specj}, \ref{fig:spech}, and \ref{fig:speck}. S:N is
80--160 in the middle of the atmospheric windows and far from strong
telluric bands (e.g.~at 1.23--1.27$\,\mu$m, or at 2.21--2.28$\,\mu$m),
40--80 where there are strong telluric absorption (e.g.~around
2.01$\,\mu$m), and 20--40 at the edges of the atmospheric windows
(e.g. above 1.34$\,\mu$m in the J band or above 2.40$\,\mu$m in the K
band).

\subsection{Near-infrared polarimetry}

We have acquired AO-assisted near-infrared polarimetric observations
of EX\,Lup in visitor mode using the NACO instrument mounted on the
UT4 telescope of the VLT on 10/11 April 2008, as part of the project
381.C-0241 (PI: \'A.~K\'osp\'al). NACO consists of the Nasmyth
Adaptive Optics System and the CONICA near-infrared camera
\citep{lenzen1998,rousset2003}. The observations were made using the
differential polarimetric imaging technique \citep[DPI; see
  e.g.][]{kuhn2001}. The basic idea of the DPI is to take the
difference of two orthogonally polarized, simultaneously acquired
images of the same object in order to remove all non-polarized
light. As the non-polarized light mainly comes from the central star,
after subtraction only the polarized light, such as the scattered
light from the circumstellar material, remains. We obtained
polarimetric images with NACO through the H filter, using a Wollaston
prism with a 2$''$ Wollaston mask to exclude overlapping beams of
orthogonal polarization, using the 27 mas pixel$^{-1}$ scale
camera. EX\,Lup was observed at four different rotator angles of
0$^{\circ}$, 45$^{\circ}$, 90$^{\circ}$ and 135$^{\circ}$. At each
angle, a three-point dithering was applied. The polarimetric
calibrators were HD\,94851 and HD\,64299. The data reduction was done
in IDL using previously developed software tools presented in details
in \citet{apai2004} and in \citet{kospal2008}. Calibration
measurements of the unpolarized standards HD\,94851 and HD\,64299
indicate that the instrumental polarization is less than 3\%.


\section{Results and analyses}

\begin{figure*}
\begin{center}
\includegraphics[angle=90,width=\textwidth]{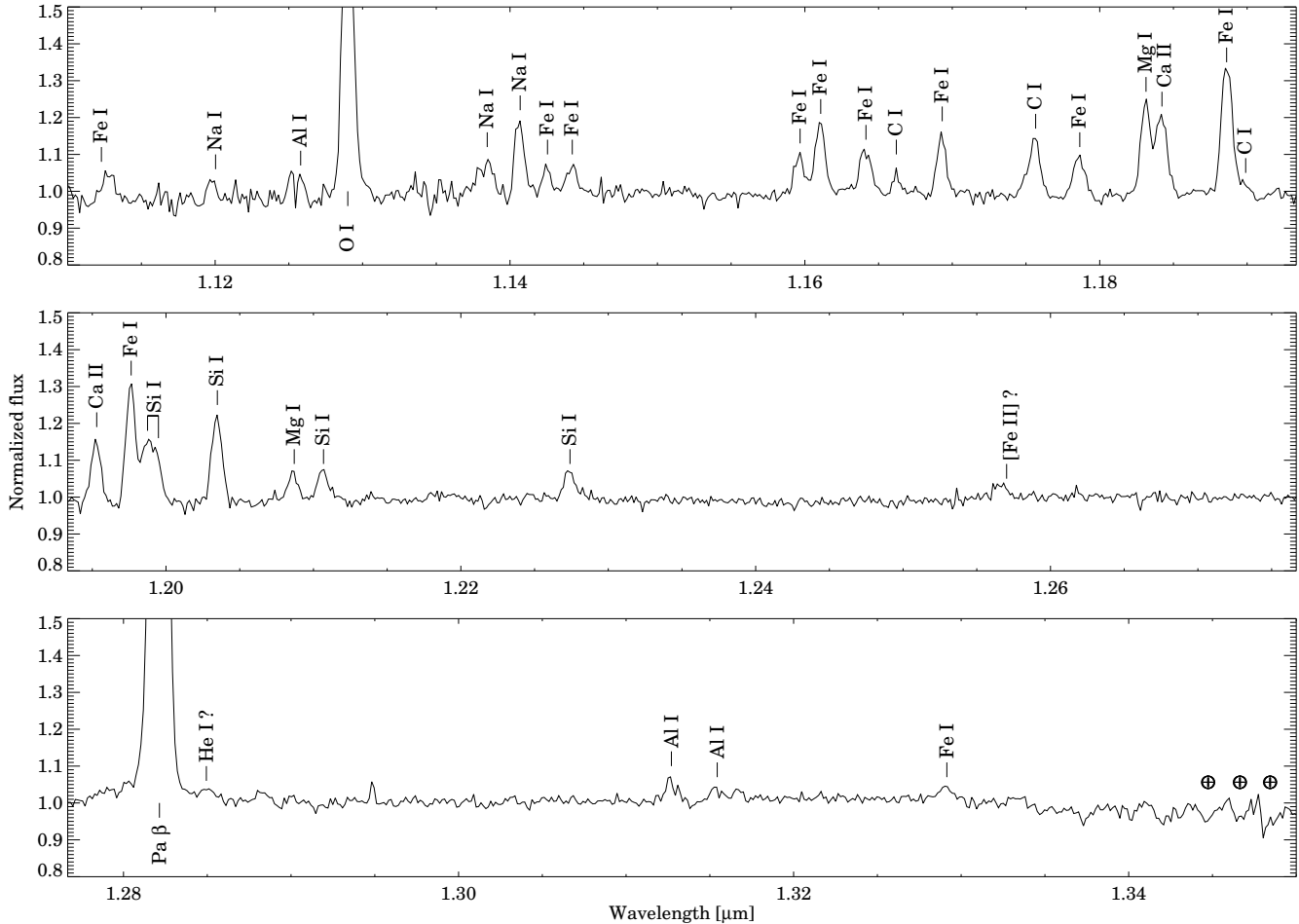}
\caption{Normalized SINFONI J band spectrum of EX\,Lup. Earth symbols
  indicate telluric absorption. \label{fig:specj}}
\end{center}
\end{figure*}

\begin{figure*}
\begin{center}
\includegraphics[angle=90,width=\textwidth]{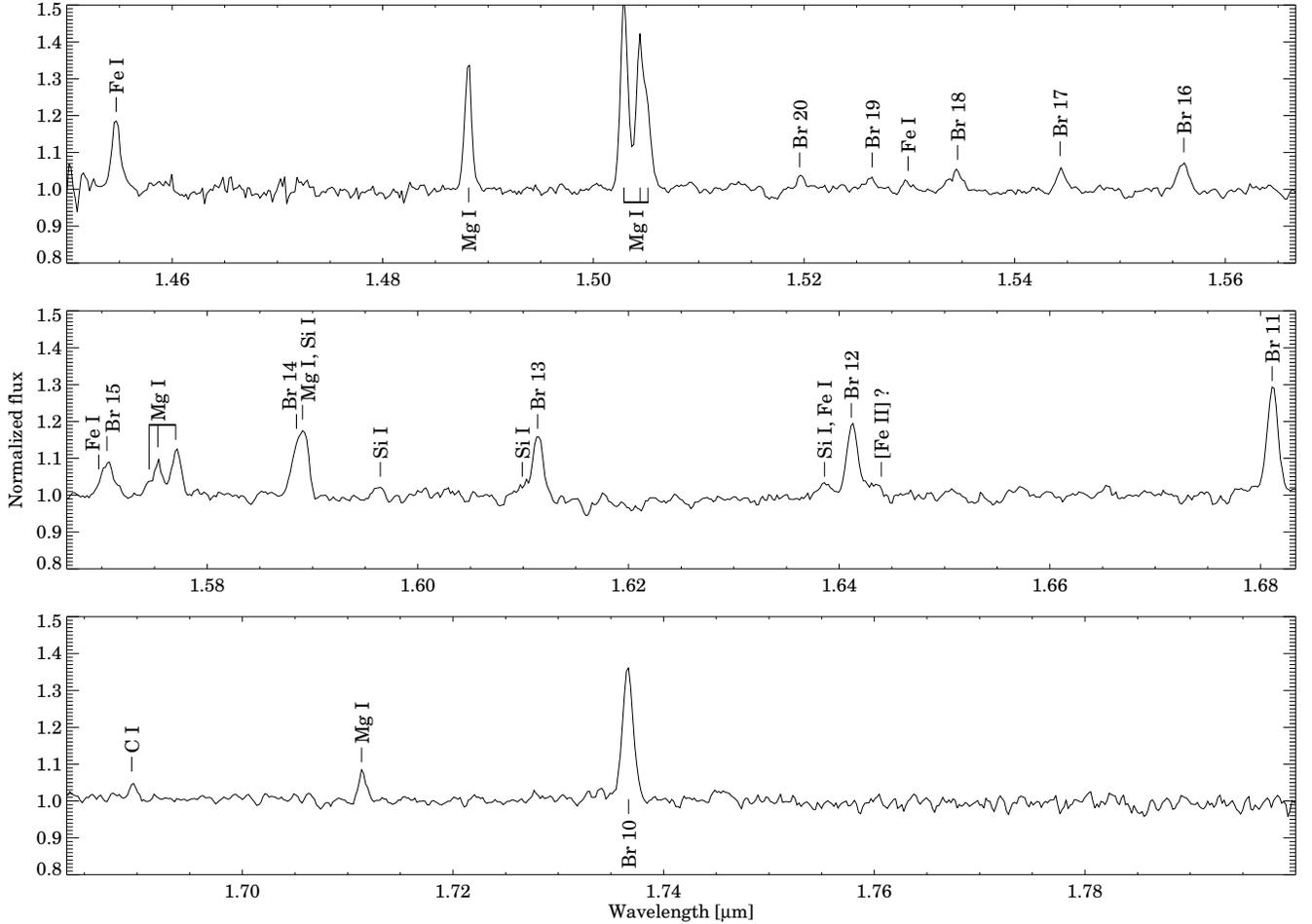}
\caption{Normalized SINFONI H-band spectrum of EX\,Lup. \label{fig:spech}}
\end{center}
\end{figure*}

\begin{figure*}
\begin{center}
\includegraphics[angle=90,width=\textwidth]{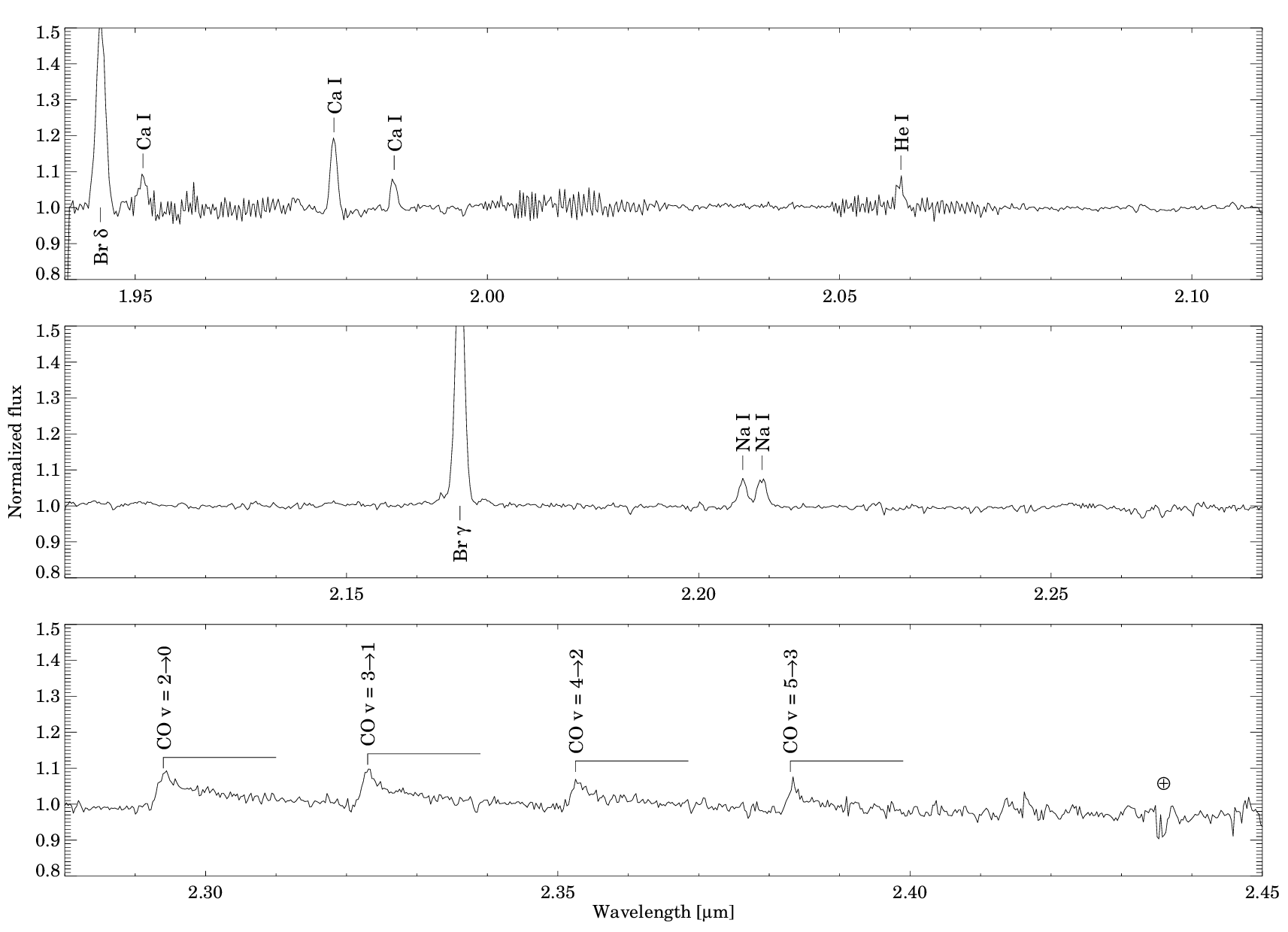}
\caption{Normalized SINFONI K-band spectrum of EX\,Lup. The earth
  symbol indicates telluric absorption. \label{fig:speck}}
\end{center}
\end{figure*}

\paragraph{Line inventory.} The near-infrared spectrum of EX\,Lup
displays many emission lines. No obvious absorption features are
visible. We identified the emission features using wavelengths from
the NIST Atomic Spectra
Database\footnote{http://physics.nist.gov/PhysRefData/ASD/lines\_form.html}
and from the Infrared Spectral Atlases of the Sun from NOAO
\citep{wallace1996}. There is no systematic difference between the
observed and tabulated wavelengths; the difference for the individual
lines being in average less than 2 pixels or 70\,km\,s$^{-1}$.

The most conspicuous features in the spectrum are the lines of atomic
hydrogen: the Pa$\,\beta$ line in the J band (the strongest feature in
the whole spectrum), the Br\,10 to 20 lines in the H band, and the
Br$\,\gamma$ and $\delta$ lines in the K band. We could identify two
He I lines: the well-known 2.0587$\,\mu$m feature, and possibly also
another line at 1.2849$\,\mu$m (close to the Pa$\,\beta$
line). Several metallic lines are present: lines of neutral Na, Mg,
Fe, Si, Ca, Al, C, and O are well visible. We could also identify two
lines of singly ionized Ca, and possibly also the 1.6440$\,\mu$m
forbidden line of singly ionized Fe. Besides the atomic lines,
molecular bands are also visible: the K band spectrum displays
pronounced CO ro-vibrational overtone bandhead emission between 2.29
and 2.40$\,\mu$m. The species of the identified lines are overplotted
in Figs.~\ref{fig:specj}, \ref{fig:spech}, and \ref{fig:speck}.

We analyzed the atomic lines in IRAF\footnote{IRAF is distributed by
  the National Optical Astronomy Observatories, which are operated by
  the Association of Universities for Research in Astronomy, Inc.,
  under cooperative agreement with the National Science Foundation.}
by fitting Gaussian profiles. We found that the full widths at half
maximum (FWHM) of the lines are consistently larger than the
instrumental FWHM, the latter being 6\,\AA{} in the middle of the J
and H bands, and 5\,\AA{} in the middle of the K band, as determined
by measuring spectrally unresolved OH sky emission lines. Thus, we
conclude that the lines are spectrally resolved, their measured FWHM
are between 8 and 12 pixel, and their deconvolved FWHM are between 110
and 190\,km\,s$^{-1}$. Table \ref{tab:spec} shows the tabulated and
observed wavelengths, as well as the equivalent widths (EWs), of the
lines identified in our EX\,Lup spectrum.

\paragraph{Spatial extent.} Since the SINFONI data cubes provide
spatially resolved information, we checked whether any line emission
originates from an extended area. We first averaged images
corresponding to wavelengths in a wavelength window of 25\,\AA{}
centered on a certain emission line (``line image''). Then we averaged
images in a 50\,\AA{} wide window on both sides of the line, at a
distance of 87.5\,\AA{} (``continuum image''). We created a
continuum-subtracted line image by simply subtracting the continuum
image from the line image. In order to do PSF-subtraction, we supposed
that the continuum emission is a point source and considered the
continuum image as a PSF template. Then, we subtracted the PSF
template from the continuum-subtracted line image. The peak of the PSF
template was scaled to the peak of the line image. If the line
emission is more extended than the continuum, the residuals should
show this extended emission. With this method, we analysed several
hydrogen and metallic lines, as well as the CO bandhead, but found no
extended emission. Thus, we can conclude that with our sensitivity,
the line emission is not more extended than the continuum emission,
and that there is no extra diffuse line emission apart from the
central star. The FWHM of the stellar PSF should give a reasonable
upper limit for the size of the emitting region. The FWHM was between
4.1 and 6.4 pixels, depending on the band and on the efficiency of the
AO correction for a particular frame. 4 pixels correspond to 50\,mas
(or approximately 8\,AU at the distance of 155\,pc). Thus, the frames
with the best AO correction give an upper limit of 8\,AU for the
diameter of the near-infrared emitting region, both for the continuum
and for the H, CO, and metallic line emission. A somewhat less
stringent upper limit can be derived from the FWHM of the broad-band H
images obtained with NACO (92\,mas or 14\,AU).

\citet{aspin2010} observed P\,Cygni profile for the H$\beta$ and the
Na D lines in the optical spectrum of EX\,Lup obtained in January
2008. The blue-shifted absorption in these lines appear around
$-$120\,km\,s$^{-1}$. \citet{goto2011} detected similar blue-shifted
absorption at $-$80\,km\,s$^{-1}$ in the profiles of the $v=1-0$
fundamental CO lines in August 2008. Outflowing material traveling at
a velocity of 80-120\,km\,s$^{-1}$ would have reached an area of
16-24\,AU in diameter during the 5 months that elapsed between the
peak of the outburst and our SINFONI observations. The fact that we do
not see emission beyond 8\,AU can probably be explained by the faint
surface brightness of the outflow. This also implies that the lines
are dominated by emission from infalling or rotating material closer
to the star than 8\,AU.

\begin{figure}
\begin{center}
\includegraphics[angle=90,width=\columnwidth]{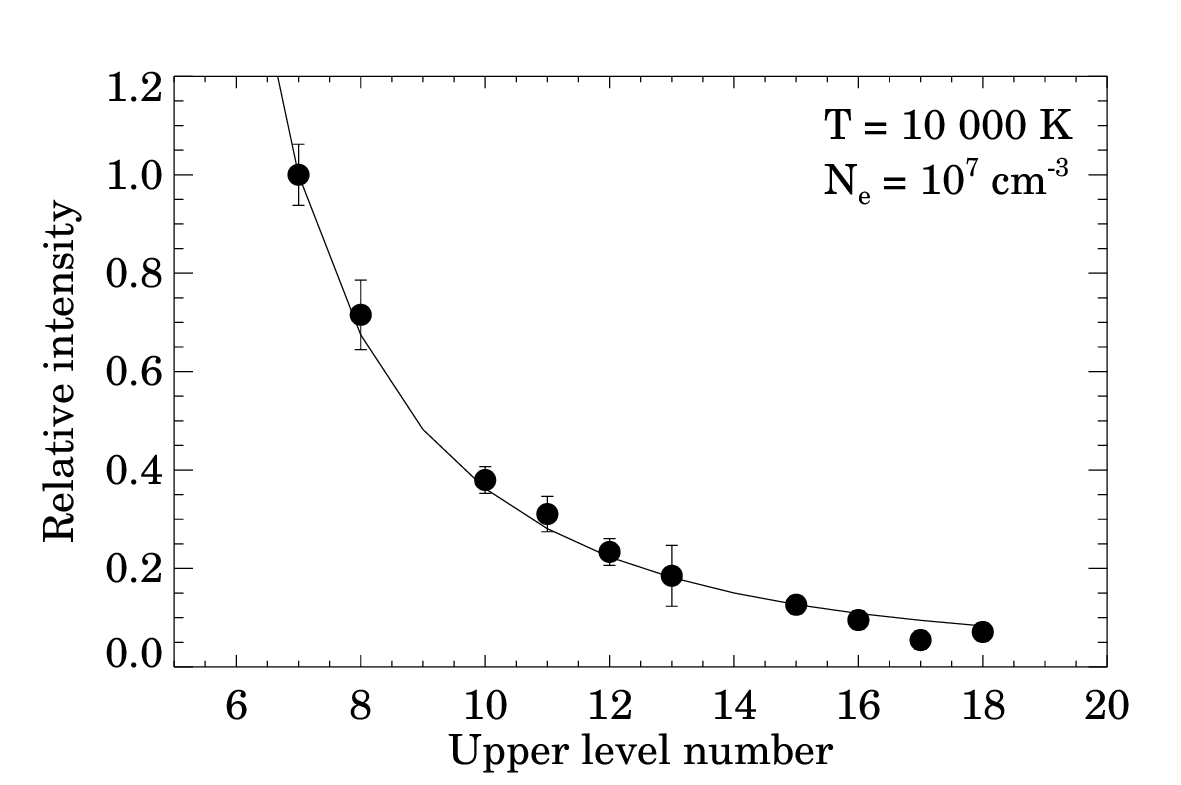}
\caption{Excitation diagram for the hydrogen Brackett series. Dots
  indicate the observed fluxes of the Brackett lines relative to
  Br\,$\gamma$. The solid line corresponds to the case B theory for a
  temperature of 10\,000\,K and electron density of 10$^7$\,cm$^{-3}$
  \citep{hummer1987}. Error bars smaller than the symbol size are not
  plotted. \label{fig:caseB}}
\end{center}
\end{figure}

\paragraph{Hydrogen lines.} Strengths and profiles of the hydrogen
lines give information about the physical conditions in the emitting
medium. In Fig.~\ref{fig:caseB} we plotted the intensities of lines in
the Brackett series relative to that of the Br\,$\gamma$ line. Here we
assumed that A$_{\rm V}=0\,$mag \citep[see e.g.][]{sipos2009,
  aspin2010}, and that the shape of the SED is flat, thus the absolute
line intensities are proportional to the line EWs. Then we compared
the observed intensity ratios to theoretical predictions corresponding
to Case B recombination \citep{baker1938, hummer1987}. Case B theory
is valid if the emitting plasma is opaque to Ly\,$\alpha$ photons but
optically thin for higher transition lines. We found that our line
fluxes are essentially consistent with the Case B theory. A model with
temperature of 10\,000\,K and electron density of N$_{\rm
  e}$=10$^7$\,cm$^{-3}$ best fitted the data points
(Fig.~\ref{fig:caseB}), however, higher electron densities with higher
temperatures (N$_{\rm e}$=10$^8$\,cm$^{-3}$, T=12\,500\,K) or lower
electron densities with lower temperatures (N$_{\rm
  e}$=10$^6$\,cm$^{-3}$, T=5000\,K) also fit within the measurement
uncertainties. The observed Pa\,$\beta$ / Br\,$\gamma$ ratio of 2.54,
however, is far from the theoretical Case B value of 6.08. Actually,
in Case B theory, any combination of temperature between 1000 and
12\,500\,K and electron density between 10$^2$ and
10$^{10}$\,cm$^{-3}$ would give ratios between 4 and 9. This indicates
that the Pa\,$\beta$ line is probably not optically thin.

\begin{figure}
\begin{center}
\includegraphics[angle=90,width=\columnwidth]{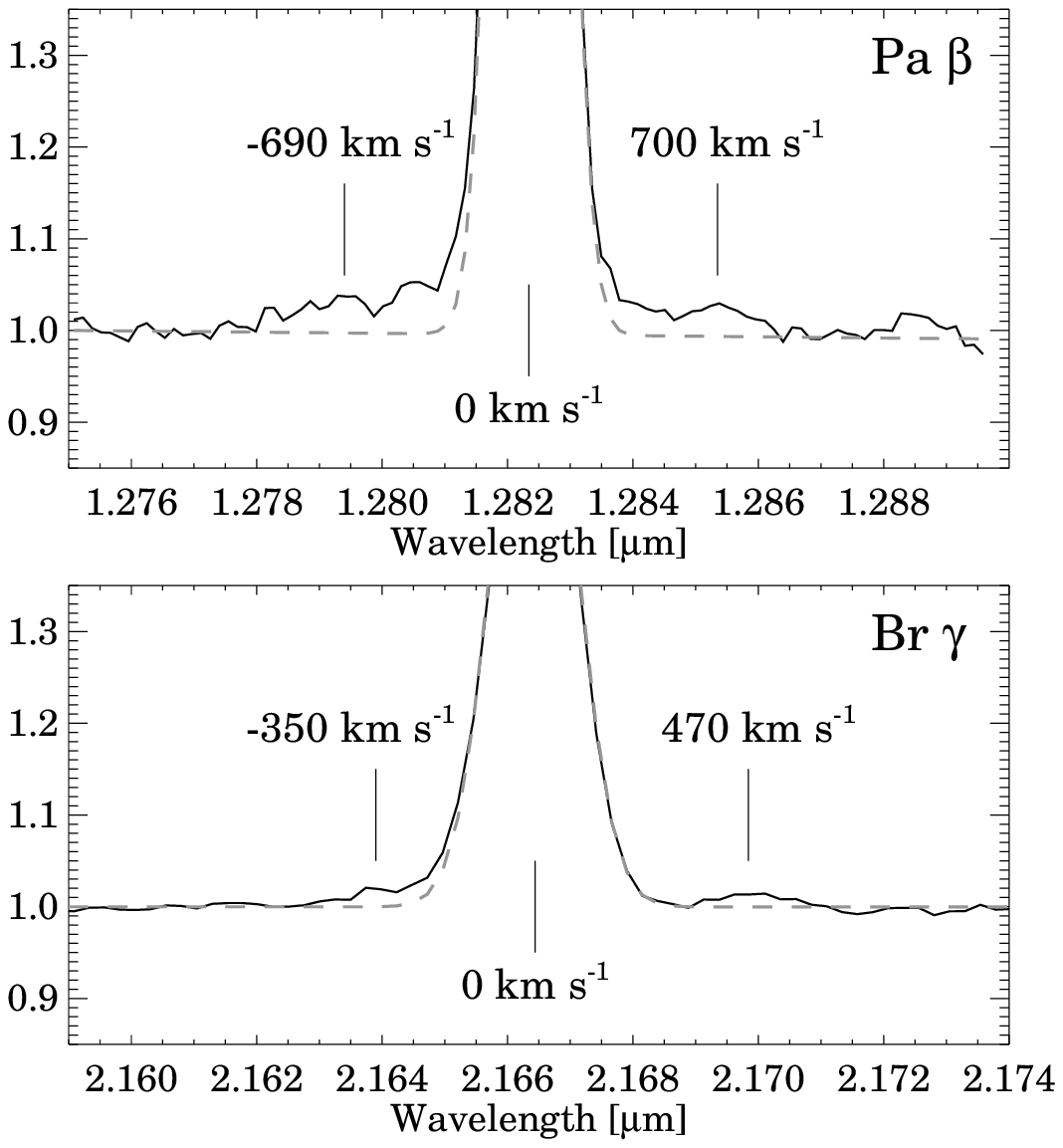}
\caption{{\it Solid black line:} profile of the Pa $\beta$ and Br
  $\gamma$ lines; {\it dashed gray line:} fitted Gaussian
  profile. \label{fig:brgamma_pabeta2}}
\end{center}
\end{figure}

As we mentioned earlier, the hydrogen lines are spectrally resolved.
Specifically, the deconvolved FWHM of the Pa\,$\beta$ line is
170$\,{\pm}\,$20\,km\,s$^{-1}$, and of the Br\,$\gamma$ line is
186$\,{\pm}\,$20\,km\,s$^{-1}$. Apart from a nearly symmetric
broadened Gaussian peak, high-velocity line wings, extending to
several hundred km\,s$^{-1}$, can also be seen
(Fig.~\ref{fig:brgamma_pabeta2}). Particularly interesting is the
shape of the Br\,$\gamma$ line wings, which is not symmetric: on the
blue side, the profile agrees with a Gaussian, and starts showing
excess from $-$230\,km\,s$^{-1}$; on the red side, the profile agrees
with a Gaussian up to 340\,km\,s$^{-1}$, and shows a pronounced bump
between 340 and 630\,km\,s$^{-1}$. This absence of excess emission in
the red side may be due to a weak inverse P\,Cygni profile, the
signature of mass infall.

Before the outburst, in 2001, EX\,Lup displayed no discernible
Br\,$\gamma$ emission \citep{sipos2009}. The spectra obtained during
outburst in April 2008 \citep{juhasz2010} and in July 2008 (this work)
both have similar EW (${\approx}\,-$13\,\AA{}) and deconvolved FWHM
(${\approx}\,$200\,km\,s$^{-1}$). Observations by \citet{aspin2010}
show that in February and in May 2008, the EW of the Br\,$\gamma$ line
was slightly larger, $-$17\,\AA{}.

\paragraph{CO bandhead.} In order to model the CO bandhead profiles,
we used a simple slab model in which we assume that the CO emission
comes from a slab of gas in local thermodynamical equilibrium (LTE)
with vibrational temperature T$_{\rm CO}$ and CO column density
N$_{\rm CO}$. The vibrational and the rotational temperature were set
to be equal. We used the equations of \citet{kraus2000}. We took the
Einstein A coefficients from \citet{chandra1996}, who calculated
coefficients up to J=140, and we took the total partition function
from \citet{goorvitch1994}. We first calculated the absorption
coefficient per CO molecule, as given by Eqn.~10 in \citet{kraus2000},
using a Gaussian CO line profile with a width of 2 km s$^{-1}$, as
suggested for the turbulent broadening by \citet{najita1996}. We only
considered temperatures between 2000\,K and 5000\,K. For LTE, we need
sufficiently high temperature and sufficiently high density so that
the vibrational levels can be collisionally excited. This requires
temperatures T$_{\rm CO}>$2000\,K and densities greater than n$_{\rm
  H} > 10^{10}$\,cm$^{-3}$; but temperatures should also be less than
5000\,K, otherwise CO molecules would dissociate \citep{scoville1980}.
Then, we convolved the absorption coefficient as a function of
wavelength with the velocity profile of a disk in Keplerian rotation
(see below). We then multiplied the absorption coefficient by the
column density of CO in order to obtain the optical depth. From the
optical depth, we calculated line intensities using the transfer
equation (Eqn.~9 in \citealt{kraus2000}).

\begin{figure*}
\begin{center}
\includegraphics[angle=90,width=\textwidth]{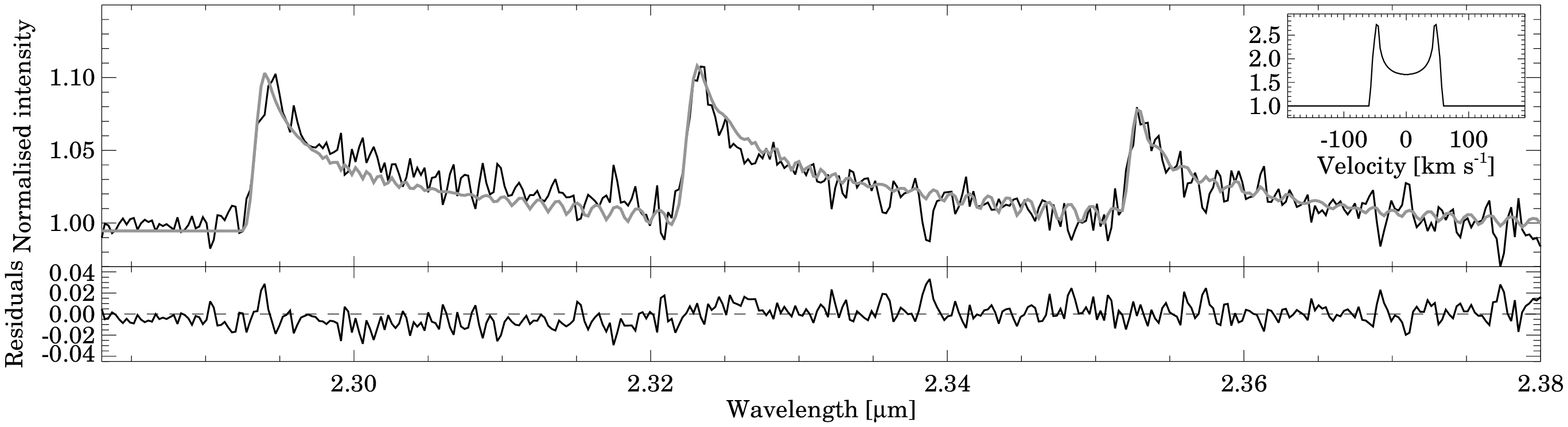}
\caption{CO overtone bandhead emission in the spectrum of EX\,Lup. The
  gray line indicates a model of 2500\,K CO gas, convolved with a line
  profile characteristic of a Keplerian disk around a 0.6\,M$_{\odot}$
  star with inner and outer radii of 0.08 and 0.13\,AU, respectively,
  and inclination of 45$^{\circ}$ (see small inset).\label{fig:co}}
\end{center}
\end{figure*}

\citet{aspin2010} needed to introduce a velocity profile of a disk in
Keplerian rotation in order to fit their observations of the CO
bandhead emission. They found that the best fit could be obtained
using a stellar mass of M$_*$=0.6\,M$_{\odot}$, inner disk radius of
r$_{\rm in}$=0.08\,AU, outer disk radius of r$_{\rm out}$=0.13\,AU,
temperature of T$_{\rm CO}$=2500\,K and optical depth $\tau_{\rm
  CO}<0.1$. We calculated the line profile emerging from such a disk,
assuming a disk inclination of i=45$^{\circ}$ and a disk radial
brightness profile proportional to r$^{-2.5}$ (same as in
\citealt{goto2011}). The resulting line profile is plotted in the
small inset in Fig.~\ref{fig:co}. We used this profile when
calculating the optical depth (see above). Finally, we smoothed our
model spectrum to the instrumental resolution of SINFONI. The result,
plotted with a gray line in Fig.~\ref{fig:co}, fits our observations
of the v=2$\rightarrow$0, v=3$\rightarrow$1, and v=4$\rightarrow$2
transitions very well. We note that a model with an outer radius of
r$_{\rm out}$=0.4\,AU, as derived by \citet{goto2011} from the
profiles of the CO fundamental emission lines of EX\,Lup, would fit
our observations equally well.

\paragraph{Metallic lines.} M-type stars display numerous Na, Ca, K,
Fe, Ti, and Si absorption lines in their near-infrared spectra, which
are often used for spectral type classification
\citep{cushing2005}. In quiescence, EX\,Lup also displayed atomic
absorption lines (S\,I at 1.199$\,\mu$m, Mg\,I at 1.577$\,\mu$m, and
Al\,I at 1.673$\,\mu$m, see \citealt{sipos2009}). These lines are most
probably photospheric in origin \citep[see also][]{herbig2001}. During
outburst, no absorption lines are visible any more, despite the fact
that, at the time of our SINFONI observation, 77\% of the J and H-band
flux, and 60\% of the K-band flux was photospheric.  These values were
obtained by supposing that the quiescent JHK photometry represent the
stellar photosphere, and that during our observations, the V$-$J,
V$-$H, and V$-$K colors were the same during our observations as in
April 2008 \citep{juhasz2010}. In outburst, the above mentioned Mg\,I
line is in emission, the lines that were in emission already in
quiescence (the Pa\,$\beta$ line and an O\,I line at 1.129$\,\mu$m)
became much stronger, and many more atomic emission lines appeared.

Most of the metallic lines visible in the SINFONI spectrum originate
from neutral atoms whose ionization potential is low (in the range of
5.1$-$8.2\,eV, cf.~13.6\,eV for hydrogen), the only exception is
oxygen (with an ionization potential of 13.6\,eV). This means that
these atoms are located in an area where hydrogen is mostly neutral,
probably shielded from most of high energy radiation from the
accretion shock region by the circumstellar disk itself. However, the
atoms cannot be located too deep in the disk, since many lines and
line ratios indicate fluorescent emission pumped by UV photons. The
Na\,I doublet at 2.206 and 2.209$\,\mu$m may be pumped by 3303\,\AA{}
continuum photons. If this is the case, then the Na\,I line at
1.141$\,\mu$m should also be present with a strength roughly twice of
those of the lines at 2.206 and 2.209$\,\mu$m, while collisional
excitation would cause a line ratio above 10 \citep{mcgregor1988}. For
EX\,Lup this ratio is $\approx$1.3, supporting the fluorescent
origin. Another fluorescent line is that of O\,I at 1.129$\,\mu$m,
which may be pumped by Ly\,$\beta$ or continuum photons \citep[][and
  references therein]{mcgregor1984}. Continuum fluorescence predicts
an O\,I 1.316$\,\mu$m / 1.129$\,\mu$m line ratio above unity. The
1.316$\,\mu$m line is practically invisible in our SINFONI spectrum,
which favors the Ly\,$\beta$ fluorescence. If this is true, another
O\,I line at 8446\,\AA{} should also be present. Indeed, this line is
well visible in our FEROS spectra of EX\,Lup obtained between April
and June 2008 \citep{sicilia-aguilar2011}. Interestingly, both the
8446\,\AA{} and the 1.129\,$\mu$m lines are visible even in
quiescence, although they are much weaker than in outburst, indicating
that a sufficient amount of Ly\,$\beta$ radiation was present even in
quiescence. The presence of Ly\,$\beta$-pumped lines of neutral oxygen
indicates the presence of dense warm regions embedded in hot ionized
plasma. The Mg\,I 1.503$\,\mu$m line may also be formed by UV
fluorescence. This requires that the Mg\,I 1.574/1.575/1.577$\,\mu$m
transitions are also present. This is true for EX\,Lup. However,
collisional excitation may also play a role here, since the Mg\,I
1.183$\,\mu$m line is also visible.

\paragraph{Shock signatures.} We searched the spectrum of EX\,Lup for
shock-excited lines. We found that the H$_2$ line at 2.1218$\,\mu$m is
not visible. The [FeII] line at 1.6440$\,\mu$m may be present, but its
unambiguous identification is difficult due to its closeness to the
strong Br\,12 line. We checked the position-velocity diagram and the
spectro-astrometric signal (see below) in order to identify the
location of the 1.6440$\,\mu$m emission. We found neither extended
emission nor positional offset compared to the continuum
position. Thus, supposing that the [FeII] line is indeed present, at
the time of our observations the shock was still very close to the
central source. This apparent lack of strong forbidden lines may be an
indication that the emission line region has a high density where
forbidden lines are quenched \citep{nisini2005}. Further monitoring of
EX\,Lup may reveal shocks as they reach less dense regions farther
from the central star, expanding into the medium around the disk.

\paragraph{Spectro-astrometry.} The SINFONI data cubes can be used for
spectro-astrometry, i.e. measuring the position of the source as a
function of wavelength. If there is extended emitting material moving
at different velocities, the source position measured at these
velocities will deviate from the position measured at continuum
wavelengths. In case the emission comes from a rotating disk, at
certain wavelengths, there will be a positive offset in the centroid,
while at other wavelengths, there will be a negative offset. Thus,
depending on whether we see the approaching or the receding part of
the disk, the measured spectro-astrometric signal as a function of
velocity will follow a typical sinusoidal pattern \citep[see
e.g.~Fig.~2 in][]{pontoppidan2008}. Infall or outflow would, in most
cases, cause a positive only or negative only signal \citep[see the
modeling by][]{eisner2010}. For certain geometries, a bipolar outflow
could cause both positive and negative offsets for different
velocities, but the centroid displacement would be more linear
\citep[see e.g.~the example of V536\,Aql in][]{whelan2004}. Binaries
may also cause a non-zero spectro-astrometric signal, but again, a
positive only or negative only signal
\citep{takami2003,brannigan2006}.

\begin{figure}
\begin{center}
\includegraphics[angle=90,width=0.5\textwidth]{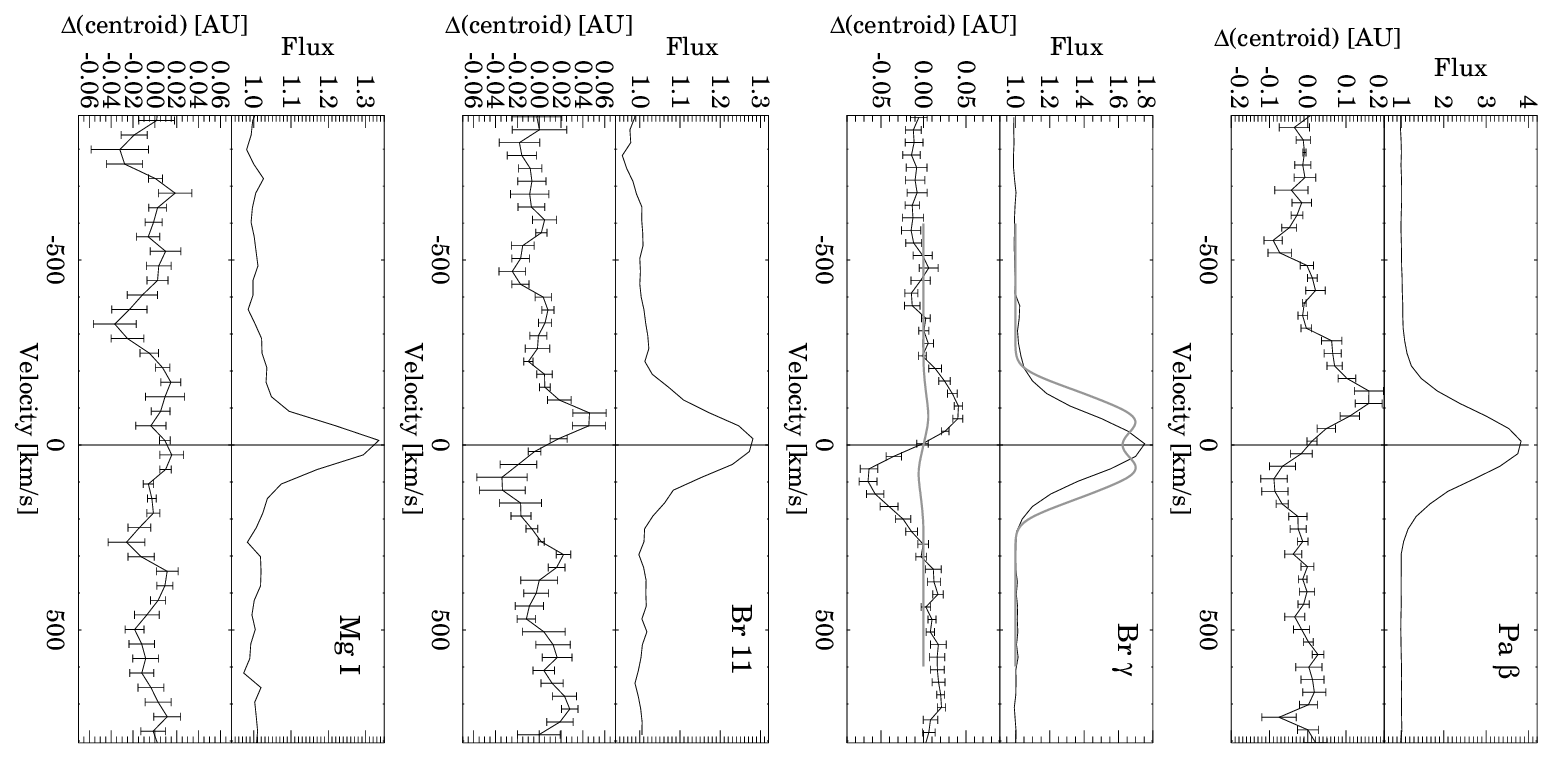}
\caption{{\it Top panels:} Profile of the Pa\,$\beta$, Br\,$\gamma$,
  Br\,11, and the 1.4882$\,\mu$m Mg\,I line (the flux is normalized to
  the continuum). {\it Bottom panels:} Displacement of the of the
  image centroid measured on the line+continuum images with respect to
  the image centroid at continuum wavelengths. The model overplotted
  with a gray line on the Br\,$\gamma$ spectro-astrometric signal is a
  Keplerian disk model.  For model parameters, and a more detailed
  discussion of the spectro-astrometric data, see
  text.\label{fig:specast}}
\end{center}
\end{figure}

We analyzed several emission lines in the spectrum of EX\,Lup with
spectro-astrometric technique. Fig.~\ref{fig:specast} shows some
examples. We found that only the hydrogen lines indicate any
spectro-astrometric signal. Metallic lines show a flat signal, which
can be either due to gas close to the star, or slow-moving gas at
larger distances. The fact that most of the metallic atoms are neutral
supports the second possibility. The spectro-astrometric signals
plotted in Fig.~\ref{fig:specast} for the Pa$\,\beta$, Br$\,\gamma$,
and Br\,11 lines are all indicating rotating material. Since SINFONI
contains no slit, our measurements do not suffer from the artifacts
usual for slit spectra, and the spectro-astrometric signal can be
calculated for any angle across the source. The results can be then
used to calculate the position angle of the rotation axis, i.e.~the
angle where the spectro-astrometric signal is constant 0. In our case,
this yields P.A. = 80$\,{\pm}\,$10$^{\circ}$, thus, the rotation axis
is almost exactly east-west oriented. Remarkably, the
spectro-astrometric signals for the hydrogen lines are all symmetric
within the measurement uncertainties, indicating that the brightness
and velocity distribution of hydrogen gas around the central star is
azimuthally symmetric.

\paragraph{Polarimetry.} Broad-band polarimetric images through the H
filter were obtained for EX\,Lup using a shorter (0.35\,s) and a
longer (2\,s) exposure time. The short exposure images show that the
distribution of the total intensity follows a Gaussian profile. The
profile of the long exposure images indicates that the inner region
(within a radius of 3 pixels) of the source is saturated, but the rest
of the profile is the same as for the short exposure images,
suggesting no extended emission. The Stokes Q and U images, as well as
the polarized intensity images show no polarized light
whatsoever. This limits the H-band degree of polarization for EX\,Lup
below the instrumental polarization (not more than 1-2\%). This result
is not surprising if we consider that only a small fraction of the
observed H-band flux is expected to be scattered light (thus
polarized); most of it is photospheric emission from the central star,
or thermal emission from the disk.

\section{Discussion}

\subsection{The origin of the CO emission}

Our model indicates that the CO bandhead emission is consistent with
emission from CO gas which has a temperature of 2500\,K and which is
located in the disk between 0.08 and 0.13\,AU from the central star.
\citet{juhasz2010} found that in outburst there is a dust-free inner
hole in the system within 0.3\,AU. Thus, the CO gas we observe here
should mostly be located in this dust-free region.

\citet{goto2011} also derived the properties and distribution of the
CO gas using the fundamental vibrational lines in the 4.6--5\,$\mu$m
wavelength range. They found that the fundamental lines are the
superposition of a narrow (FWHM=50\,km s$^{-1}$) and a broad (FWZI=150
km s$^{-1}$) component. The broad component comes from gas orbiting
the central star at 0.04--0.4\,AU, and has a vibrational temperature
between 1800 and 3000\,K. These parameters are very similar to those
we obtained from the overtone bandhead features. Thus the gas emitting
the overtone and the fundamental lines are co-located and are very
likely physically the same material.

The CO bandhead feature in EX\,Lup displays significant temporal
variability: it was in absorption in quiescence
\citep{sipos2009,herbig2001}, very strong emission in February 2008,
weak emission in May 2008 \citep{aspin2010}, and also weak emission in
July 2008 (this work). \citet{lorenzetti2009} observed the variability
of the CO bandhead feature in several other EXors. They claim that CO
absorption is associated with quiescent periods, when the accretion
rate is low and the stellar photosphere (of an M-type dwarf) is
visible. According to them, CO emission is associated with more active
periods, when the accretion rate is higher, UV radiation is higher as
well, thus the CO gas in the inner disk is heated and is producing CO
bandhead emission. The general behavior of the CO bandhead feature in
EX\,Lup is consistent with this picture.

There are several scenarios described in the literature to explain the
origin of CO bandhead emission in YSOs. Based on observations of
low-mass YSOs indicating significant CO bandhead emission variability
on timescales as short as a few days, \citet{biscaya1997} considered
the disk, stellar/disk wind, and funnel flows as the origin of CO
emission. Although our observations are consistent with the inner disk
origin (Fig.~\ref{fig:co}), we cannot exclude the funnel flow origin
either: the figure in \citet{martin1997} suggests that funnel flows
can produce CO bandhead profiles very similar to those we
observed. \citet{goto2011} argued against the funnel flow origin based
on the lack of systematic redward absorption in the fundamental CO
lines. However, one can imagine a situation when -- due to the
inclination of the system -- the funnel flows do not cross the
line-of-sight toward the hot spot on the stellar surface, thus,
absorption is not expected to be seen. We note that CO can survive
without dissociation only in the outer parts of the funnel flows,
close to the inner edge of the disk, thus, there is no essential
physical difference between a funnel flow and an inner disk origin.

In order to test whether the heating of the CO gas is related to
accretion, at different epochs we correlated the strength of the CO
bandhead feature with the visual brightness of the star (which is
supposed to be accretion luminosity) and with the Br\,$\gamma$ line
flux (another usual accretion rate tracer). Using the spline-smoothed
version of the light curve presented in \citet{juhasz2010}, we
estimated a visual brightness of 9.5\,mag in February, 10.2\,mag in
May, and 10.2\,mag in July 2008. Assuming that the V$-$K color of the
source was similar at these epochs to the value measured in April 2008
(2.05\,mag, \citealt{juhasz2010}), the Br\,$\gamma$ line fluxes
decreased by a factor of 1.9 between February and May, and a factor of
2.3 between February and July. We calculated the EWs of the CO
bandhead by integrating the spectra between 2.2925 and 2.2975$\,\mu$m,
and obtained $-$21\,\AA{} for February, $-$4\,\AA{} for May, and
$-$3.3\,\AA{} for July 2008. Assuming again a constant V$-$K color,
the flux of the CO bandhead decreased by a factor of 10 between
February and May, and a factor of 12 between February and July. Thus,
the strength of the CO bandhead feature, the optical brightness of the
star, and the strength of the Br\,$\gamma$ feature seems to be
changing in the same direction, although not exactly with the same
rate. Consequently, our results indicate that both heating by
irradiation and heating by accretion may have a role in the excitation
of the CO bandhead. We note that the broad component observed by
\citet{goto2011} in the fundamental emission lines was also fading
with time, but due to different observing epochs, we cannot directly
compare the results for time evolution of the fundamental and the
overtone lines.

\subsection{The origin of metallic lines}

The lack of spectro-astrometric signal for the metallic lines and the
fact that most of the metallic lines are from neutral atoms indicate
that these lines are originating from regions farther from the central
star than the hydrogen lines. Some of the lines are fluorescently
excited by UV photons, indicating that at least part of the atoms are
subject to direct stellar radiation, which condition is probably
fulfilled at least on the surface of the inner gas disk. Thus, the
metallic atoms may be partially co-located with the CO
gas. Interestingly, the 2.206 and 2.209$\,\mu$m Na doublet shows a
similar time evolution to that of the CO
bandhead. \citet{lorenzetti2009} noticed that it follows the same
behavior as the CO bandhead in several other EXors: either both are in
emission or both are in absorption. They interpreted this result as a
common origin: both the Na and the CO feature is originated in the
stellar photosphere when in absorption, an in the inner disk when in
emission. Our results for EX\,Lup is consistent with this scenario:
both features were in absorption in quiescence \citep{sipos2009}, and
were in emission in outburst \citep[][and this work]{aspin2010}.

\subsection{The origin of the hydrogen emission}

Hydrogen emission in YSOs is generally believed to be originating in
disks, funnel flows or stellar/disk winds
\citep[e.g.][]{najita1996b,kraus2010}. In the following, we check
whether any of these scenarios are consistent with our results on
EX\,Lup. The excitation diagram in Fig.~\ref{fig:caseB} indicates that
the hydrogen is hotter than the CO gas, suggesting the hydrogen gas is
closer to the star than 0.04-0.08\,AU. Following
\citet{pontoppidan2008} and \citet{eisner2010}, we modeled the
Br\,$\gamma$ line using this scenario and adopting a simple disk
model. We supposed that all continuum arises from (and only from) the
central star, and all the Br\,$\gamma$ emission originates from (and
only from) the disk. We assumed that the velocity field in the disk is
Keplerian around a 0.6\,M$_{\odot}$-mass central star, while the
brightness profile is a power law. With these assumptions, we
calculated the line-of-sight component of the velocity, and made
images corresponding to each SINFONI velocity channel. We added a
central point source with a wavelength-independent brightness, and
blurred the image by convolving it with a Gaussian PSF. We measured
the spectro-astrometric signal on these synthetic images the same way
as we did for the real observations. We used a disk extending from the
stellar surface (0.007\,AU) to 0.04\,AU, inclination of 45$^{\circ}$,
and a radial brightness profile proportional to r$^{-2.5}$. The
resulting line profile and spectro-astrometric signal are overplotted
with gray lines in Fig.~\ref{fig:specast}. The results indicate that
emission from such a Keplerian disk cannot significantly contribute to
the observed Br\,$\gamma$ line, because the emerging line profile
would be double-peaked and the model spectro-astrometric signal
indicates that high-velocity gas is closer to the star than observed.

We calculated different disk models by changing the disk inclination,
disk inner and outer radii, and power law exponent of the disk
brightness profile. We could not find a model that fits the observed
spectro-astrometric signal. This analysis seems to exclude the
equatorial boundary layer as the main path of accretion onto the star,
a model often invoked for FU\,Orionis-type outbursts \citep{hk96},
especially since the innermost parts of a boundary layer is
sub-Keplerian. Although the boundary layer, or strong turbulence could
explain the single-peaked line profile, they cannot reproduce the
high-velocity high-amplitude spectro-astrometric signal.

Our results suggest that hydrogen gas is orbiting the central star but
its velocity profile is not Keplerian, in a sense that there is
high-velocity gas farther from the star. This points to a funnel flow
or disk wind origin, where the material is launched from the inner
disk along the magnetic field lines with high velocities. This may
also help to explain why the hydrogen gas is hotter than the CO. The
wind scenario is supported by the P\,Cygni profiles observed in
several optical lines by \citet{aspin2010}. Supposing that the disk
wind scenario is true for EX\,Lup, it should be optically thin,
otherwise the excitation diagram in Fig.~\ref{fig:caseB} would show
flatter line ratios \citep{lorenzetti2009}.

\section{Conclusions}

In this paper we performed a near-infrared spectroscopic and
spectro-astrometric study of EX\,Lup during its most extreme outburst
in 2008. Our main conclusions are the following:
\begin{itemize}
\item The CO bandhead emission comes from an inner, dust-free
  region. This area is identical with the broad component area
  identified by \citet{goto2011}, from where the CO fundamental
  emission comes from. The CO gas is rotating around the central star,
  either at the inner edge of the disk or in the outer parts of funnel
  flows.
\item EX\,Lup also displays numerous neutral atomic lines, which
  probably come from an area similar to that of the CO gas. At least
  part of this area is subject to direct UV photons, as evidenced by
  the presence of fluorescent emission lines.
\item A very conspicuous feature in the spectrum of EX\,Lup is the
  Brackett series, whose excitation diagram indicates that the
  hydrogen emission is optically thin. Spectro-astrometric analysis of
  these lines suggests that the hydrogen emission is probably not
  coming from an equatorial boundary layer; a funnel flow or disk wind
  origin is more likely.
\end{itemize}

Based on our findings, we can attempt to reconstruct the geometrical
and kinematic structure of the circumstellar material within the inner
few tenths of AU in the EX\,Lup system during the outburst. The dust
disk has an inner radius of 0.2-0.3\,AU
\citep{sipos2009,juhasz2010}. Within this area, there is an optically
thin gas disk, whose temperature is a few thousand K, as indicated by
the CO and neutral metal emission lines. This area certainly also
contains hydrogen gas, but this component is not visible in the
spectro-astrometric signal due to its low velocity. We see evidence
for high temperature ($\approx$10\,000\,K), high velocity
($\approx$100\,km\,s$^{-1}$) hydrogen gas in the system, which is not
located in the equatorial plane. Part of this hydrogen probably falls
onto the stellar surface along magnetic funnel flows. Indeed, based on
X-ray and UV data, \citet{grosso2010} reported on the presence of
accretion shocks and accretion hot spots on the stellar surface of
EX\,Lup. Some of the UV photons thus generated must reach the disk
surface and produce the fluorescent emission we observe in certain
sodium, magnesium and oxygen lines. The UV radiation may also trigger
chemical changes in the disk. Part of the hydrogen gas does not fall
onto the stellar surface but leaves the system in the form of a hot
wind, as evidenced by the P\,Cygni profile of the optical hydrogen
lines. Considering all these arguments, the emerging picture is
broadly consistent with that of the standard magnetospheric accretion
model usually assumed for normally accreting T\,Tauri stars
(e.g.~\citealt{bouvier2007}).

Several models are described in the literature to explain the
increased accretion in young eruptive stars. Our results may place
constraints on the applicability of these models for the case of
EX\,Lup. Our results do not indicate the presence of a fully ionized
ring of material during the eruption of EX\,Lup, which seems to
contradict the thermal instability model of \citet{hk96} and
\citet{bell1994}. No stellar or planetary companions to EX\,Lup have
been found so far \citep[][and references therein]{sipos2009}, neither
do our spectro-astrometric analysis suggests the presence of one. This
makes outburst models involving perturbation by a close companion
unlikely \citep{bonnell1992}. The model of \citet{vorobyov2010}
involves gravitational instability and fragmentation in the outer
disk, and the infall of these fragments onto the star. The modest disk
mass, and the fact that our spectro-astrometric observations indicate
an azimuthally symmetric mass distribution in the inner disk reasons
against this model.

Recently, \citet{dangelo2010} proposed that accretion onto a strongly
magnetic protostar is inherently episodic if the disk is truncated
close to the corotation radius. In their model, the magnetic field
initially truncates the disk outside the corotation radius, thus
accretion onto the star is inhibited. As gas in the inner regions of
the disk piles up, material is pushed inside the corotation radius,
and the accumulated material is accreted onto the star until the
reservoir is depleted, and the inner radius of the disk is again
outside the corotation radius. \citet{sipos2009} reported a $v \sin i$
of 4.4$\,{\pm}\,$2\,km\,s$^{-1}$ for EX\,Lup, which translates into a
rotation period of about 13 days (using an inclination of 45$^{\circ}$
and stellar radius of 1.6\,R$_{\odot}$), and a corotation radius of
about 0.3\,AU (using a stellar mass of 0.6\,M$_{\odot}$). This value
is close to the radius of the dust-free zone. Our results show that
during outburst, gas is present in this area, potentially channeled
along magnetic funnel flows. Thus, the model of \citet{dangelo2010}
may be applicable for EX\,Lup.


\acknowledgments

The results published in this paper are based on data collected at the
European Southern Observatory in the frame of the programs 281.C-5031
and 381.C-0243. We thank the referee, Colin Aspin, whose suggestions
helped to impove this paper. \'A.~K.~would like to thank Uma Gorti and
Arjan Bik for useful discussions about near-infrared spectroscopy. The
research of \'A.~K.~is supported by the Nederlands Organization for
Scientific Research. Zs.~R. has been supported in part by the DAAD-PPP
mobility grant P-M\"OB/841/ and by the ``Lend\"ulet'' Young Researcher
Program of the Hungarian Academy of Sciences.

{\it Facilities:} \facility{VLT (SINFONI)}, \facility{VLT (NACO)}.


\bibliographystyle{apj}
\bibliography{paper}{}

\clearpage

\begin{deluxetable}{ccccc}
\tablewidth{0pt}
\tablecaption{Spectral features in the near-infrared spectrum of
  EX\,Lup.\label{tab:spec}}
\tablehead{
\colhead{Species} & \colhead{$\lambda_{\rm tab}$ [$\mu$m]} &
\colhead{$\lambda_{\rm obs}$ [$\mu$m]} & \colhead{EW [\AA]} &
\colhead{Comment}
}
\startdata
Fe I              & 1.1123 & 1.1131 & $-$0.73 & \\
Na I              & 1.1200 & 1.1200 & $-$0.44 & \\
Al I              & 1.1258 & 1.1261 & $-$0.52 & \\
O I               & 1.1290 & 1.1293 & $-$7.97 & \\
Na I              & 1.1385 & 1.1386 & $-$1.19 & \\
Na I              & 1.1407 & 1.1409 & $-$1.60 & \\
Fe I              & 1.1425 & 1.1428 & $-$0.45 & \\
Fe I              & 1.1442 & 1.1445 & $-$0.61 & \\
Fe I              & 1.1597 & 1.1599 & $-$0.93 & \\
Fe I              & 1.1610 & 1.1613 & $-$1.69 & \\
Fe I              & 1.1641 & 1.1644 & $-$1.23 & \\
C I               & 1.1662 & 1.1665 & $-$0.38 & \\
Fe I              & 1.1693 & 1.1696 & $-$1.30 & \\
C I               & 1.1757 & 1.1758 & $-$1.57 & \\
Fe I              & 1.1786 & 1.1789 & $-$1.16 & \\
Mg I              & 1.1831 & 1.1834 & $-$2.02 & \\
Ca II             & 1.1842 & 1.1845 & $-$2.17 & \\
Fe I              & 1.1886 & 1.1889 & $-$3.30 & \\
C I               & 1.1899 & 1.1901 & $-$0.30 & \\
Ca II             & 1.1953 & 1.1955 & $-$1.58 & \\
Fe I              & 1.1976 & 1.1979 & $-$2.65 & \\
Si I              & 1.1987 & 1.1990 & $-$1.43 & \\
Si I              & 1.1995 & 1.1997 & $-$1.11 & \\
Si I              & 1.2035 & 1.2037 & $-$2.12 & \\
Mg I              & 1.2087 & 1.2089 & $-$0.71 & \\
Si I              & 1.2107 & 1.2109 & $-$0.94 & \\
Si I              & 1.2274 & 1.2276 & $-$0.82 & \\
H I               & 1.2821 & 1.2824 & $-$29.34& Pa$\beta$ \\
He I              & 1.2849 & 1.2854 & $-$0.20 & \\
Al I              & 1.3127 & 1.3129 & $-$0.51 & \\
Al I              & 1.3154 & 1.3157 & $-$0.35 & \\
Fe I              & 1.3292 & 1.3293 & $-$0.41 & \\
\hline
Fe I              & 1.4547 & 1.4550 & $-$1.82 & \\
Mg I              & 1.4882 & 1.4884 & $-$2.75 & \\
Mg I              & 1.5029 & 1.5032 & $-$4.49 & \\
Mg I              & 1.5044 & 1.5047 & $-$3.39 & \\
Mg I              & 1.5052 & 1.5055 & $-$1.76 & \\
H I               & 1.5195 & 1.5200 & $-$0.35 & Br 20 \\
H I               & 1.5265 & 1.5267 & $-$0.36 & Br 19 \\
Fe I              & 1.5299 & 1.5300 & $-$0.24 & \\
H I               & 1.5346 & 1.5347 & $-$0.82 & Br 18 \\
H I               & 1.5443 & 1.5447 & $-$0.63 & Br 17 \\
H I               & 1.5561 & 1.5562 & $-$1.11 & Br 16 \\
Fe I, H I         & 1.5697,1.5705 & 1.5708 & $-$1.47 & Fe I + Br 15 blend \\
Mg I              & 1.5745 & 1.5747 & $-$0.38 & \\
Mg I              & 1.5753 & 1.5757 & $-$0.99 & \\
Mg I              & 1.5770 & 1.5774 & $-$1.51 & \\
H I, Mg I, Si I   & 1.5885, 1.5891, 1.5893 & 1.5892 & $-$3.24 & Br 14 + Mg I + Si I blend \\
Si I              & 1.5964 & 1.5967 & $-$0.47 & \\
Si I              & 1.6099 & 1.6102 & $-$0.53 & \\
H I               & 1.6114 & 1.6117 & $-$2.17 & Br 13 \\
Si I, Fe I        & 1.6386, 1.6387 & 1.6390 & $-$0.78 & blend \\
H I               & 1.6412 & 1.6415 & $-$2.72 & \\
$[$Fe II$]$       & 1.6440 & 1.6436 & $-$0.74 & \\
H I               & 1.6811 & 1.6815 & $-$3.56 & Br 11 \\
C I               & 1.6895 & 1.6900 & $-$0.39 & \\
Mg I              & 1.7113 & 1.7117 & $-$0.72 & \\
H I               & 1.7367 & 1.7369 & $-$4.36 & Br 10 \\
\hline
H I               & 1.9451 & 1.9455 & $-$8.87 & Br $\delta$ \\
Ca I              & 1.9511 & 1.9515 & $-$1.20 &  \\
Ca I              & 1.9782 & 1.9786 & $-$2.43 &  \\
Ca I              & 1.9868 & 1.9871 & $-$0.98 &  \\
He I              & 2.0587 & 2.0589 & $-$0.90 &  \\
H I               & 2.1661 & 2.1665 & $-$11.54& Br $\gamma$ \\
Na I              & 2.2062 & 2.2066 & $-$1.13 &  \\
Na I              & 2.2090 & 2.2093 & $-$1.32 &  
\enddata
\end{deluxetable}

\end{document}